\documentclass[11pt,twoside]{article}

\usepackage{asp2004}
\usepackage{graphicx}

\begin{document}
\title{Spectropolarimetry of solar prominences}   
\author{R. Ramelli, M. Bianda}   
\affil{Istituto Ricerche Solari Locarno, CH-6605 Locarno Monti, Switzerland} 
\author{J. Trujillo Bueno, L. Merenda}
\affil{Instituto de Astrof\'isica de Canarias, E-38205 La Laguna, Tenerife, Spain}
\author{J. O. Stenflo}
\affil{Institute of Astronomy, ETH-Zentrum, 8092 Z\"urich, Switzerland
  {\rm \&} Faculty of Mathematics and Science, University of Zurich, 8057 Z\"urich, Switzerland}

\begin{abstract} 
A large set of high precision full-Stokes spectropolarimetric
observations of prominences in He-D$_3$, H$\alpha$ and H$\beta$ lines has been
recorded with the ZIMPOL polarimeter at the Gregory-Coud\'e Telescope
in Locarno. The observational technique allows us to obtain
measurements free from seeing induced spurious effects. The
instrumental polarization is well under control and taken into account
in the data analysis. We present our observational results for each of
the above-mentioned lines. Of particular interest is that most of our
H$\alpha$ measurements show 
anti-symmetric $V$~profiles
that are a characteristic signature
of the Zeeman effect in the prominence plasma. A Stokes inversion
technique based on the quantum theory of the Hanle and Zeeman
effects is being applied on observed Stokes profiles in the He-D$_3$ line
in order to obtain information on the magnetic field vector that
confines the prominence plasma.
\end{abstract}

\section{Introduction}

Using the same effective technique applied to spicule observations as 
described in another paper
in these proceedings \citep{Ramellispic},
it is possible also to investigate the magnetic fields present in
prominences, via spectropolarimetric measurements.
Recent examples of observations of the He-D$_3$ multiplet are reported
also by \citet{casini03} and \citet{2005smp..conf..215R}, while \citet{trujillo02} and \citet{merenda06}
have analyzed observations of the He {\sc i} 10830 \AA\ multiplet.

A few years ago, we started at the Istituto Ricerche Solari
Locarno (IRSOL) an extended observational 
project 
on prominences in He-D$_3$, H$\alpha$
and H$\beta$ with
the Zurich Imaging Polarimeter (ZIMPOL) \citep{gandorfer04}.
A preliminary physical interpretation of our observations based on suitable 
inversion techniques has been applied to the He-D$_3$ data in order
to obtain information on the magnetic field vectors involved.
The interpretation of H$\alpha$ and H$\beta$ Stokes profiles
is however postponed to a future work.

\section{The observations}
The observations were performed with the Gregory-Coud\'e
Telescope at IRSOL with the same technique applied to spicules
described in \citet{Ramellispic}.
The measurements were performed between May 2003 and
June 2005 in different positions and prominences: 49 measurements in the He-D$_3$ line, 29 in H$\alpha$ and 9 in H$\beta$.
Particular attention has been given to the correction for scattered
light as described in \citet{2005smp..conf..215R}.

\section{Results}
\begin{figure}[!h]
\includegraphics[width=.47\linewidth,angle=0]{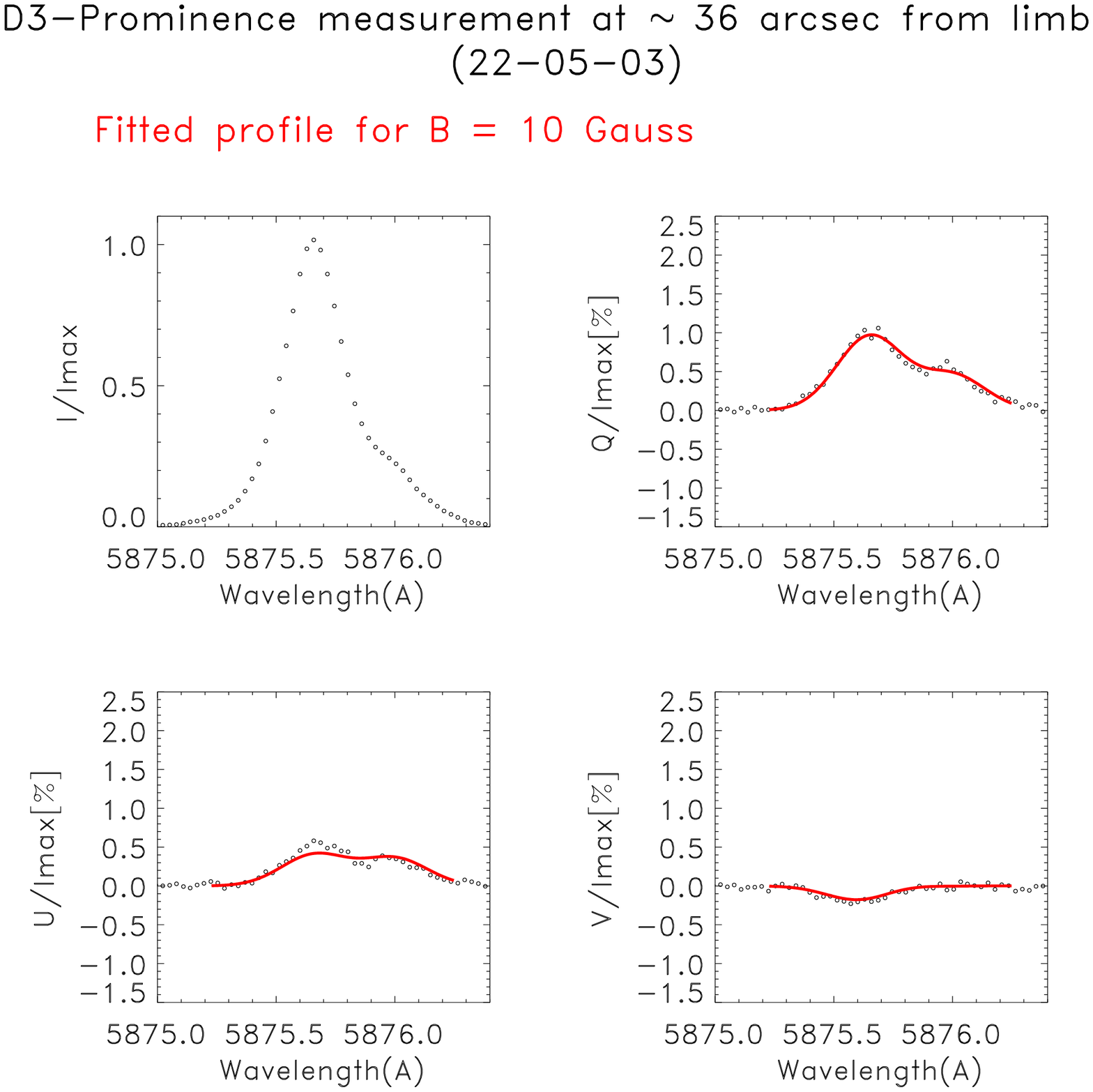}
\hfill
\includegraphics[width=.47\linewidth,angle=0]{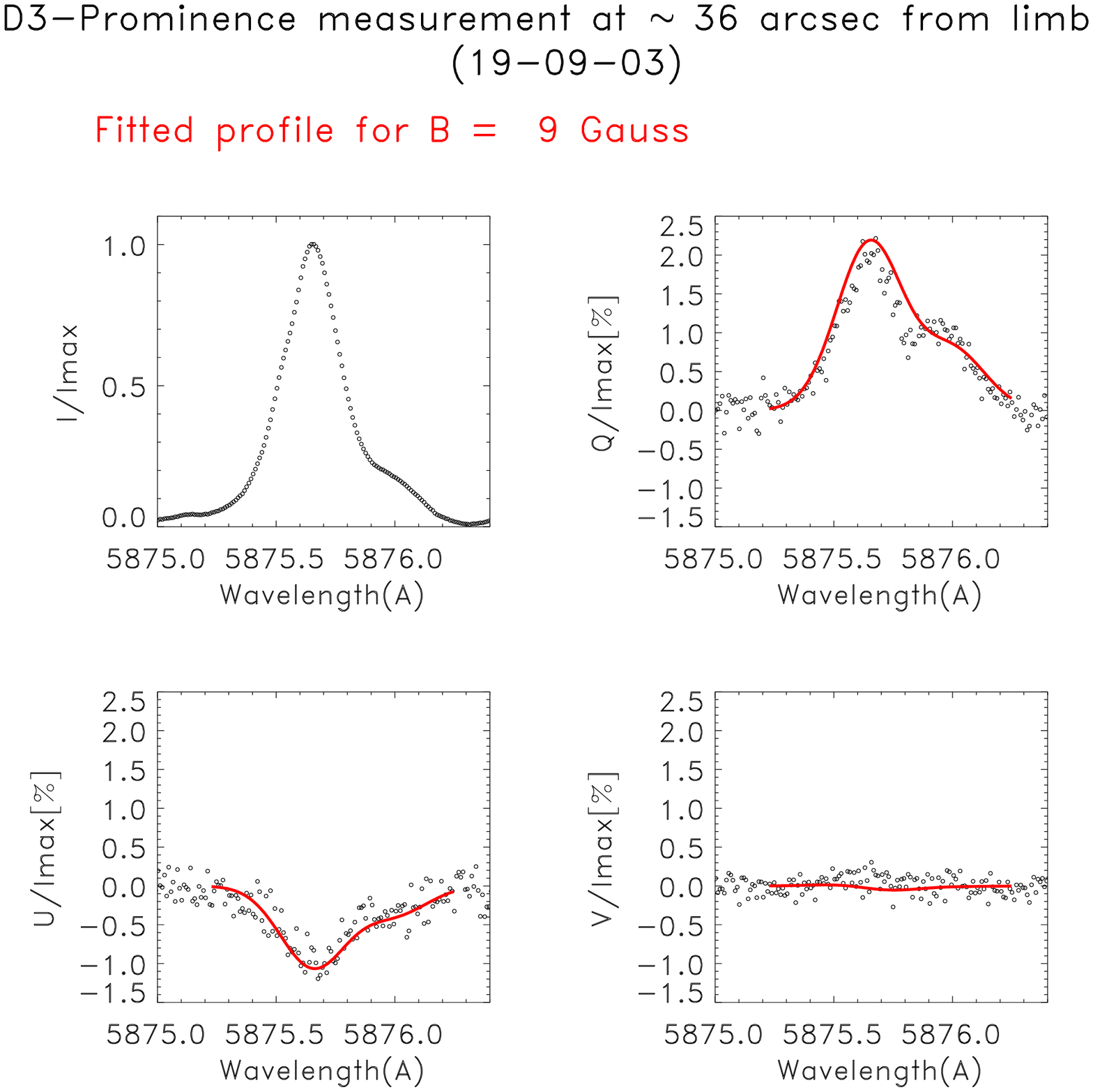}
\caption{\label{prof-prom00} Two examples of He-D$_3$ Stokes profiles obtained
from prominence observations with fitted theoretical profiles (continuous line).}
\end{figure}

A preliminary inversion of the observed Stokes profiles
has been applied to part of our He-D$_3$
observations. In this first analysis, the Stokes profiles used for 
the inversion are obtained integrating several arcseconds
along the spatial region observed through the spectrograph slit
where, in the center of 
the He-D$_3$ line, we have more than half of the maximum intensity.
Two examples of He-D$_3$ Stokes profiles with inversions are shown in 
Fig.~\ref{prof-prom00}. We usually find magnetic fields of the order of 10 gauss.

\begin{figure}[!h]
\includegraphics[angle=90,width=.47\linewidth]{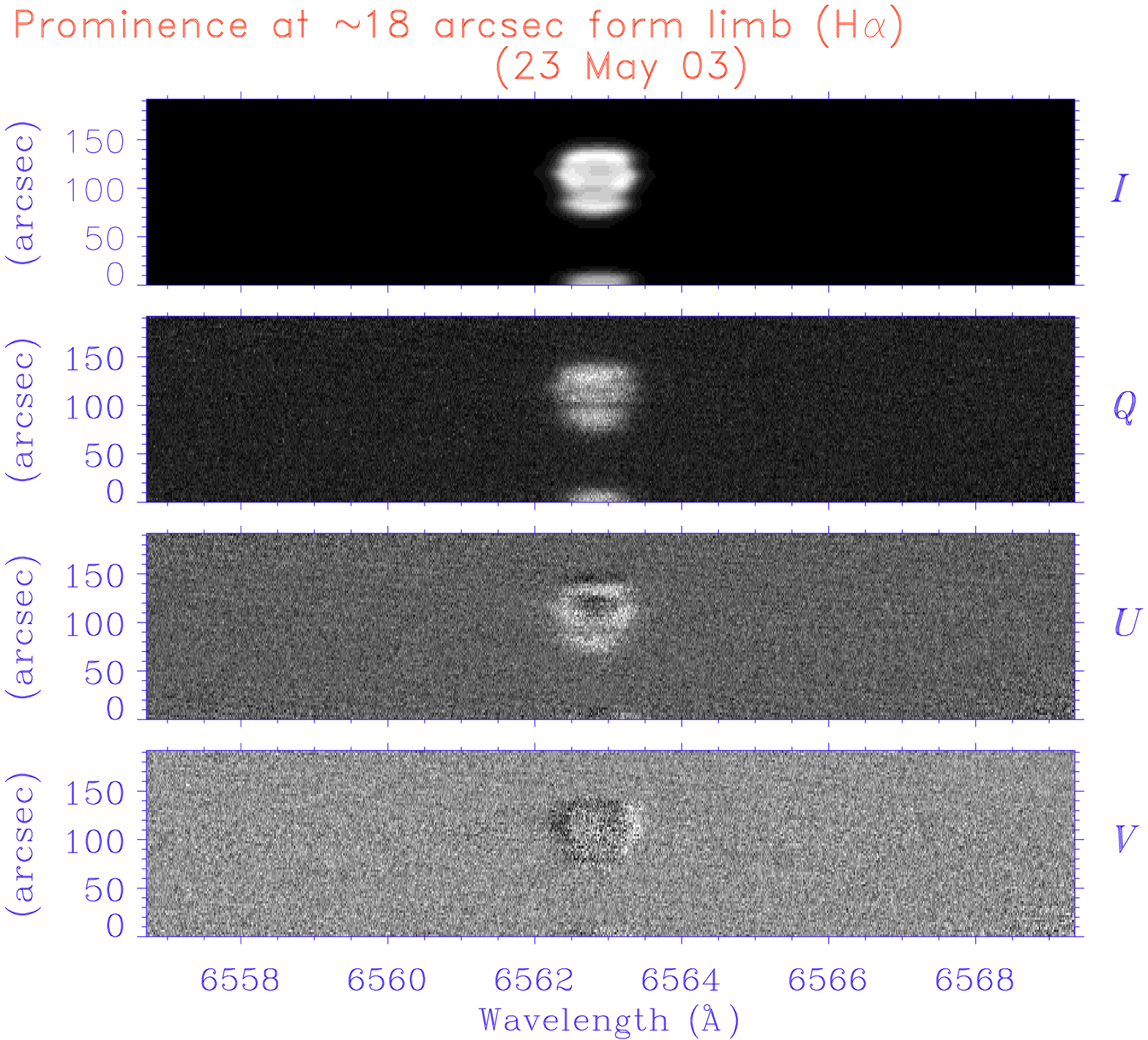}
\hfill
\includegraphics[angle=90,width=.47\linewidth]{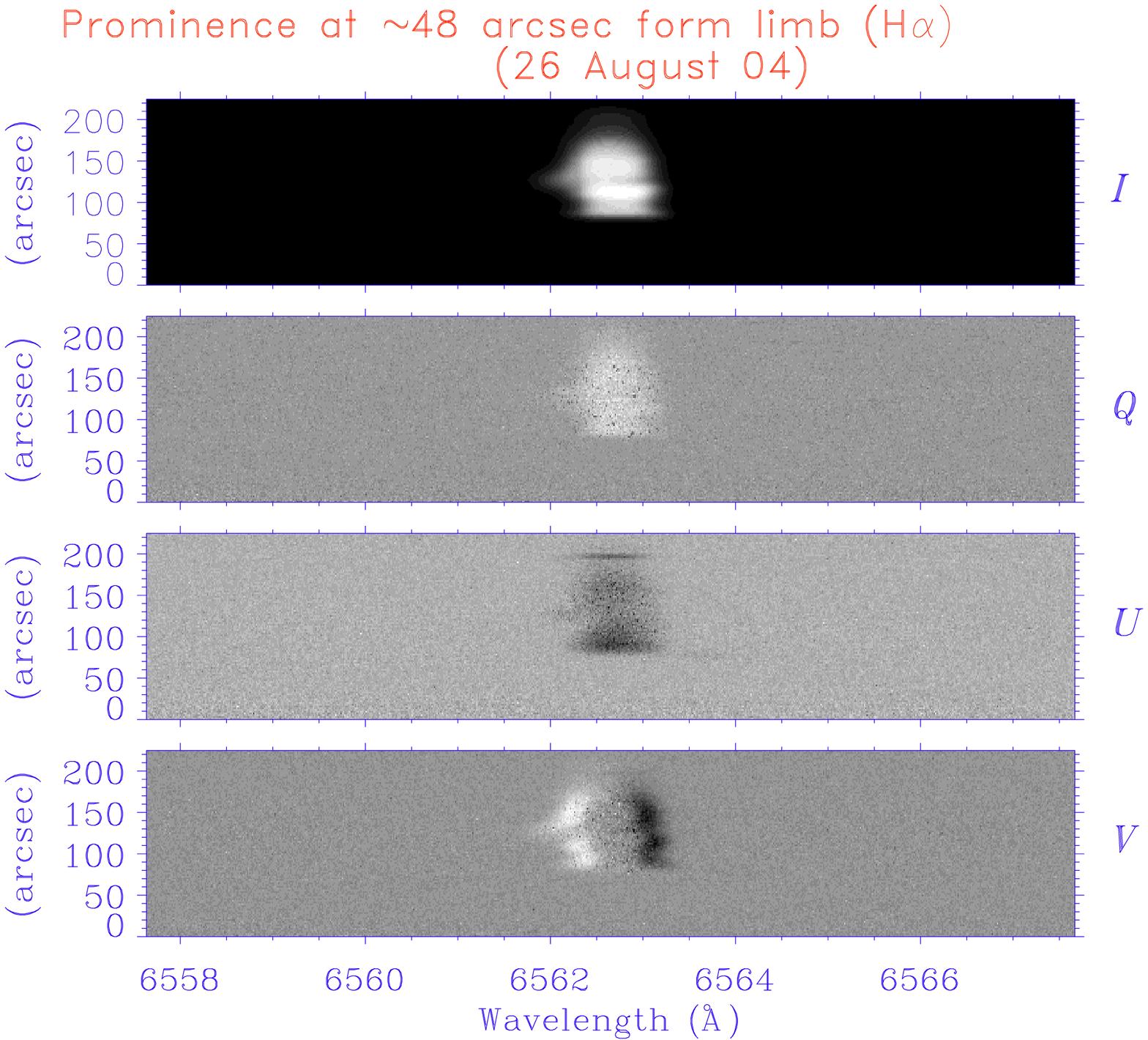}
\caption{\label{6563-030523measb}Two examples of H$\alpha$ measurements.}
\end{figure}

\begin{figure}[!h]
\includegraphics[width=.47\linewidth,angle=0]{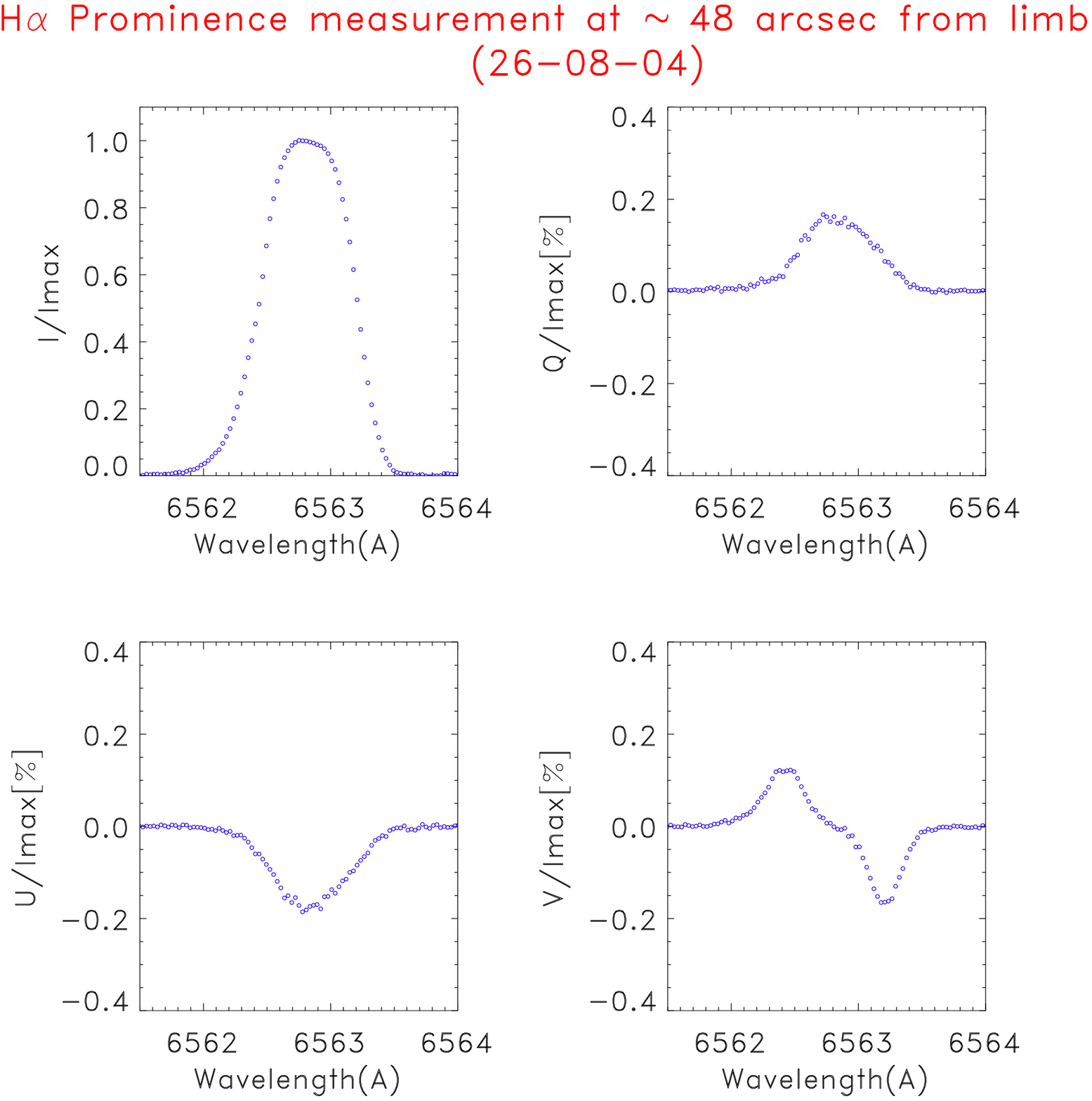}
\hfill
\includegraphics[width=.47\linewidth,angle=0]{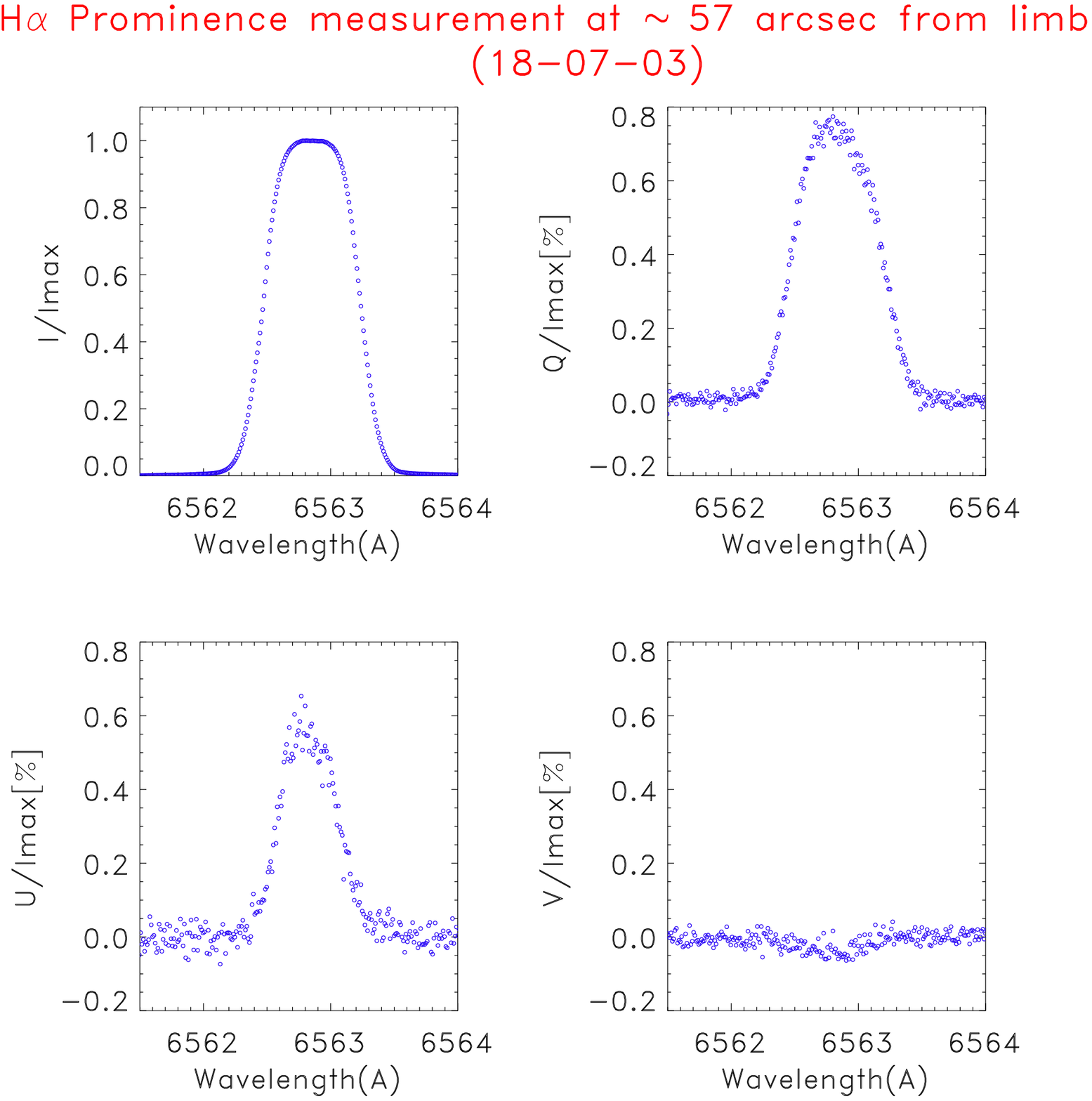}
\caption{\label{haprof}Two examples of H$\alpha$ Stokes profiles. In the first
  example left we see a typical antisymmetric Zeeman like
  structure in Stokes $V$. 
  The second example right is the only case where we find a symmetric
  structure in Stokes $V$. The amplitude in this case is only a few $10^{-4}$.}
\end{figure}

\begin{figure}[!h]
\includegraphics[angle=90,width=.47\linewidth]{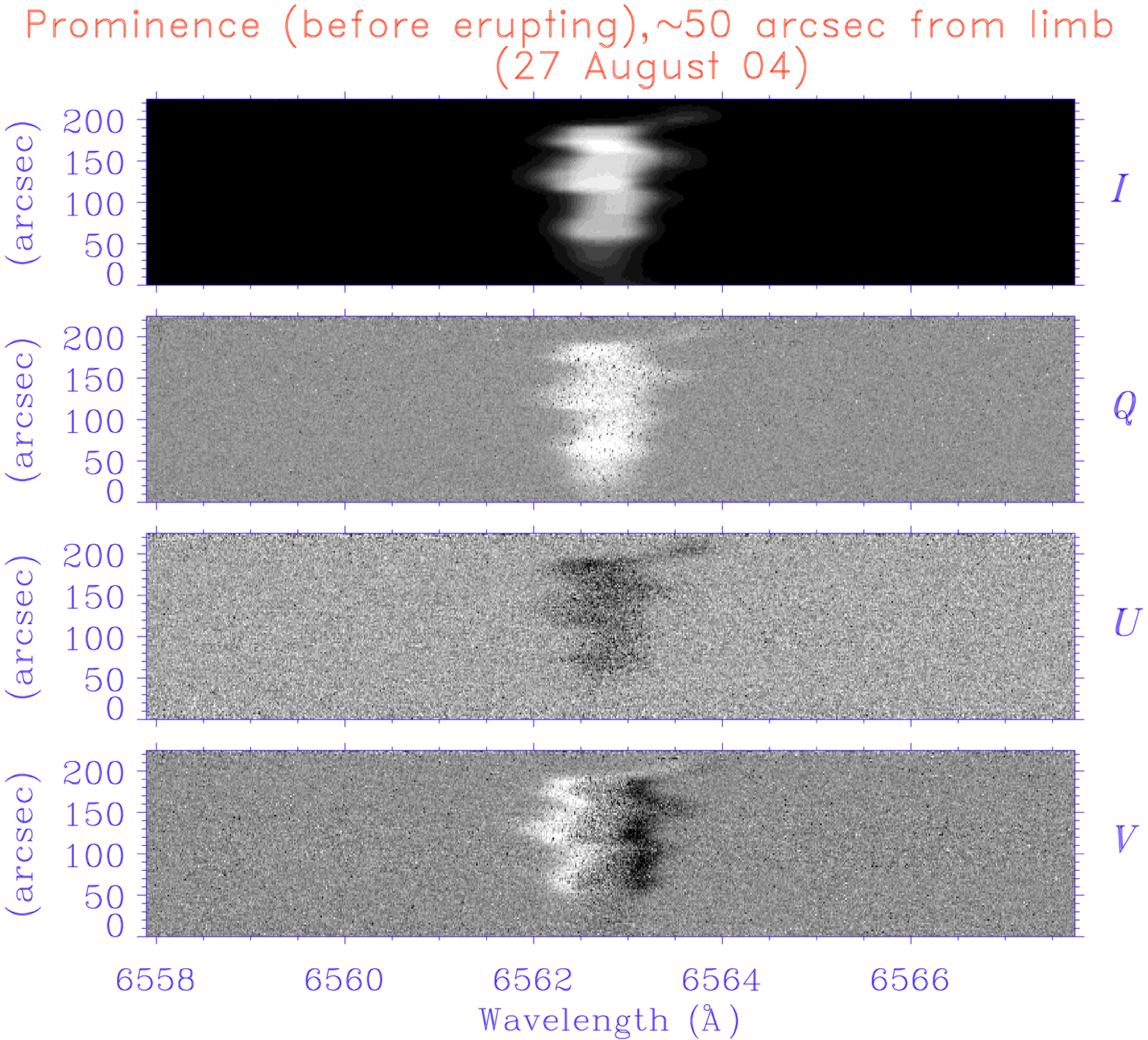}
\hfill
\includegraphics[angle=90,width=.47\linewidth]{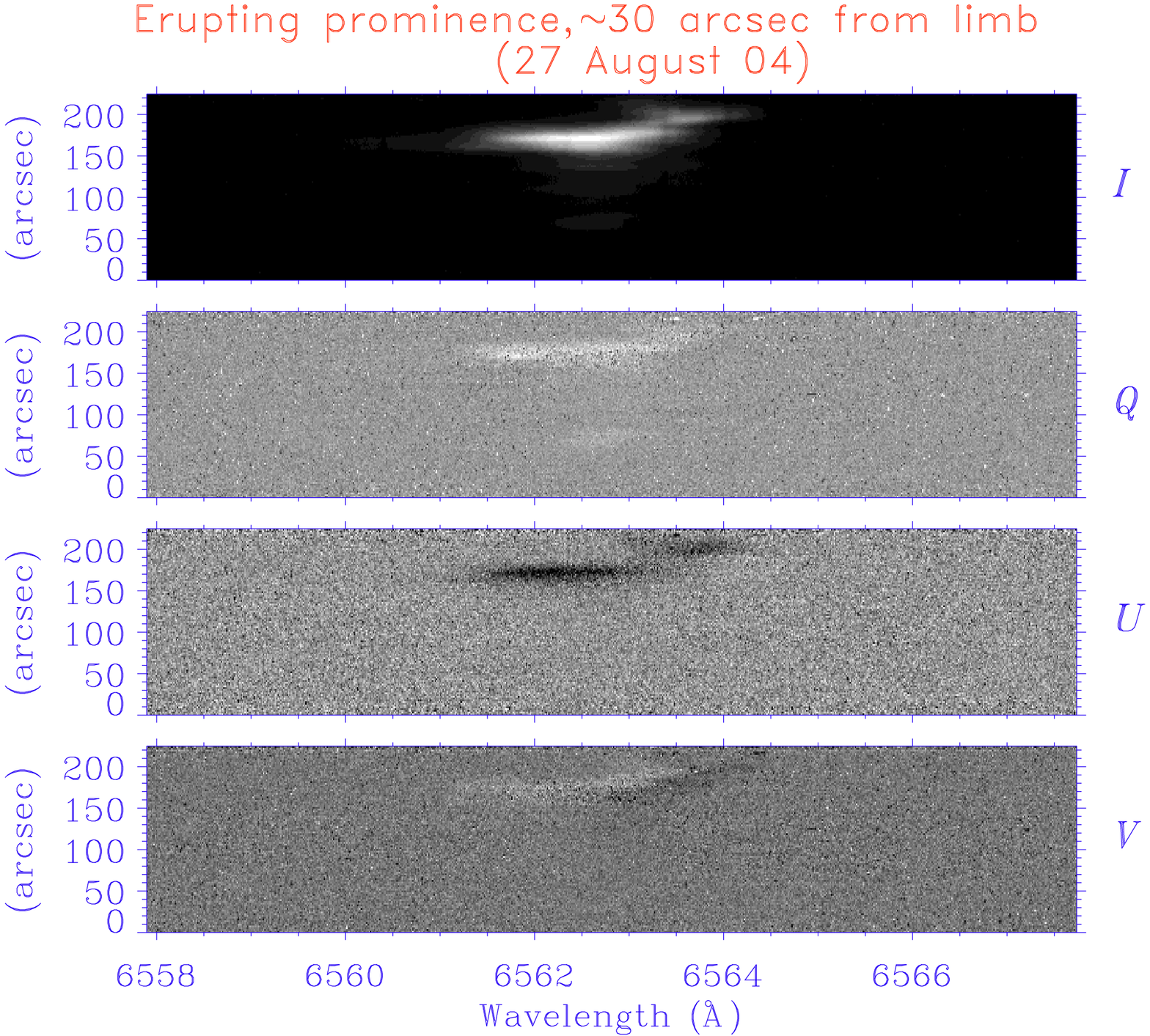}
\caption{\label{6563-040827pr6563m1} Example of H$\alpha$
measurements. {\itshape Left:\/} Prominence before erupting 
{\itshape Right:\/} Prominence during eruption }
\end{figure}

Analyzing our H$\alpha$ measurements we find that
the Stokes $V$~profiles show
usually a typical antisymmetric Zeeman like structure (e.g. 
Figs.~\ref{6563-030523measb}, \ref{haprof}, left panel, and \ref{6563-040827pr6563m1}). 
In the only example we have found a symmetric 
Stokes $V$~profile (Fig.~\ref{haprof}, right panel) 
the amplitude was very small (a few $10^{-4}$). 
Therefore, our observational results are different from those presented
by \citet{lopez05}, whose observed $V$~profiles show generally a larger symmetric
signature often dominating the antisymmetric Zeeman effect signal.
It is also interesting
to note that it is quite common to observe
self-absorption in the center of the H$\alpha$ line,
which suggests radiative transfer effects (see, e.g.,
Fig.~\ref{6563-030523measb}).

\begin{figure}[!h]
\includegraphics[angle=0,width=.47\linewidth]{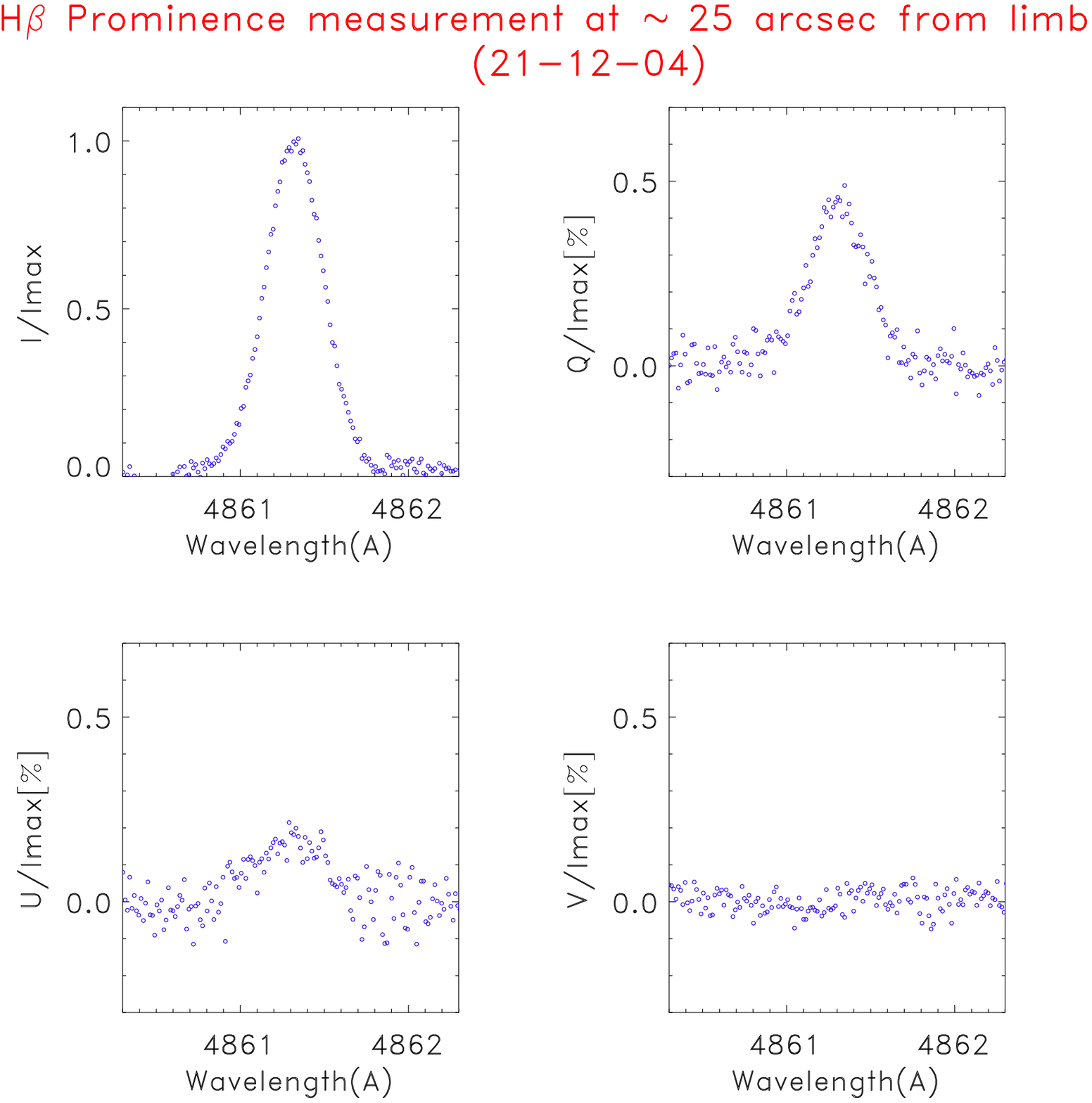}
\hfill
\includegraphics[angle=0,width=.47\linewidth]{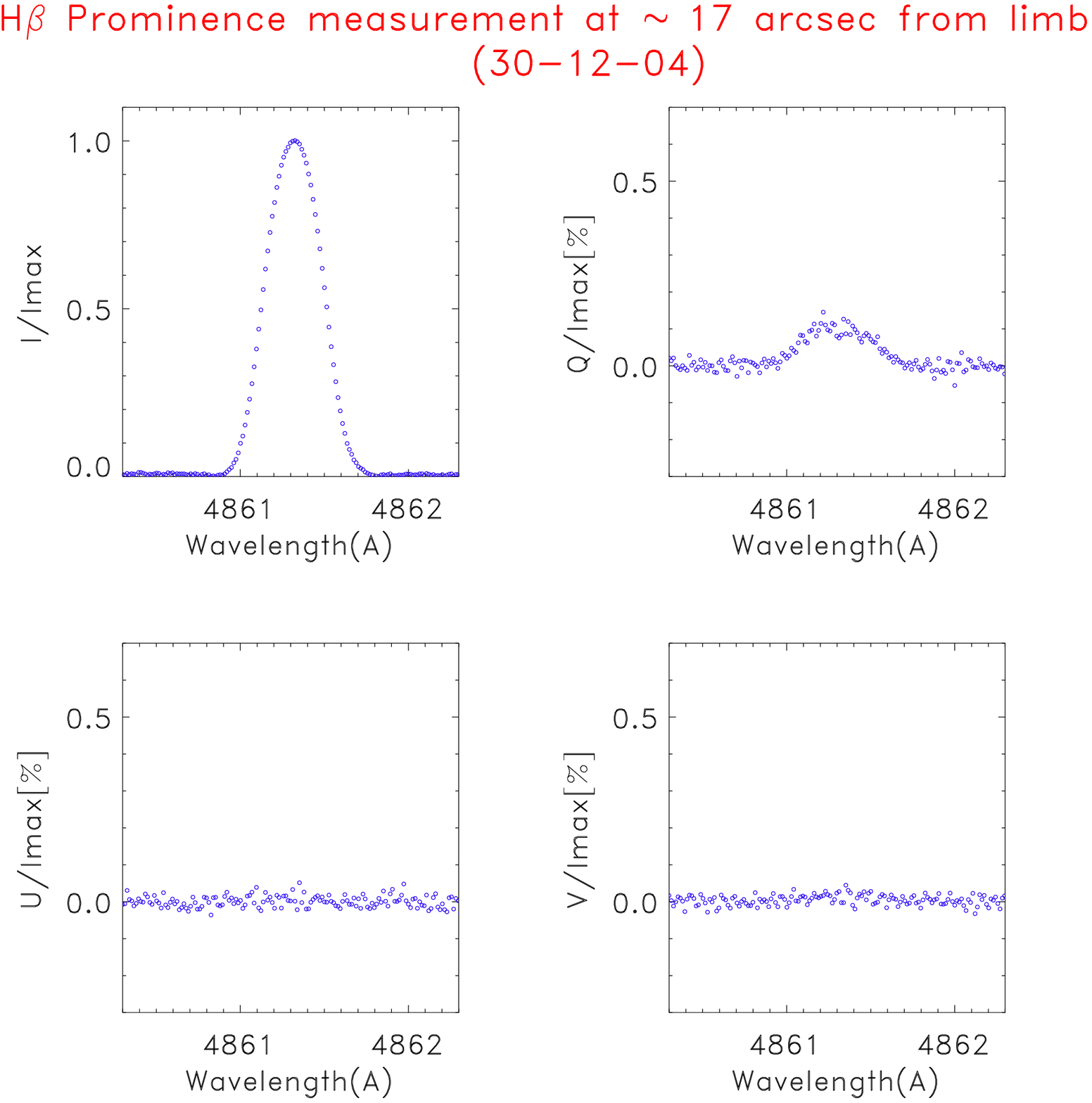}
\caption{\label{hbprof-prom05} Two examples of H$\beta$ recordings of
  Stokes profiles.}
\end{figure}

In the H$\beta$ measurements smaller polarization 
signals than in H$\alpha$ are observed. In the linear polarization
profiles, the largest amplitudes found are around
0.4\%. No signal above the noise level could be detected in the 
circular polarization.
Examples of H$\beta$ Stokes profiles are shown in Fig.~\ref{hbprof-prom05}.

\section{Conclusion}

Using ZIMPOL at the GCT telescope in Locarno it has been possible to obtain
a large set of high quality full Stokes
spectropolarimetric measurements of prominences in the 
He-D$_3$, H$\alpha$ and H$\beta$ lines.
A preliminary inversion has been applied on part of 
the He-D$_3$ measurements.

\acknowledgements 
We are grateful for the financial support that has been provided by
the canton of Ticino, the city of Locarno, ETH Zurich and the
Fondazione Carlo e Albina Cavargna.
This work has been also partially supported by the Spanish Ministerio de
Educaci\'on y Ciencia through project AYA2004-05792 and by the European
Solar Magnetism Network.

\end{document}